\documentstyle[11pt,aas2pp4,tighten,flushrt]{article}

\def\ni{\noindent}
\def\.{\mathaccent 95}
\def\a{\alpha}
\def\be{\beta}
\def\bomega{{\bf \omega}}
\def\ga{\gamma}

\def\ep{\epsilon}

\def\Om{\Omega}

\def\frac#1#2{{\textstyle{{#1}\over {#2}}}}
\def\ni{\noindent}
\def\lsim{\mathrel{\rlap{\lower4pt\hbox{\hskip1pt$\sim$}}
    \raise1pt\hbox{$<$}}}
\def\gsim{\mathrel{\rlap{\lower4pt\hbox{\hskip1pt$\sim$}}
    \raise1pt\hbox{$>$}}}
\def\sqr#1#2{{\vcenter{\vbox{\hrule height.#2pt
         \hbox{\vrule width.#2pt height#1pt \kern#1pt
         \vrule width.#2pt}
         \hrule height.#2pt}}}}

\font\gkvec=cmmib10                         %for boldface lowercase
\def\bomega{\hbox{{\gkvec\char33}}}     

\def\bw{\bar{\omega}}
\def\bv{\bar v}
\def\ts{\times}
\def\lb{\langle}
\def\rb{\rangle}
\def\curl{\nabla {\ts}}

\def\bbv{\bar {\bf v}}
\def\bfvp{{\bf v}'}
\def\bfjp{{\bf j}'}
\def\bfv{{\bf v}}
\def\bfw{{\bomega}}
\def\bbw{\bar{\bf \bomega}}
\def\bfwp{{\bomega}'}
\def\bfbp{{\bf b}'}

\def\bbb{\bar{ \bf b}}
\def\bfb{{\bf b}}
\def\nb{\nabla}
\def\curl{\nb\ts}

\def\b0{b'^{(0)}}
\def\v0{v'^{(0)}}
\def\w0{\omega'^{(0)}}
\def\bb0{\bfbp^{(0)}}
\def\bv0{\bfvp^{(0)}}
\def\bw0{\bfwp^{(0)}}
\def\bj0{\bfjp^{(0)}}

\def\bbm{\bar {\bf m}}
\def\ni{\noindent}

\begin{document}

%\draft

%\twocolumn[\hsize\textwidth\columnwidth\hsize\csname @twocolumnfalse\endcsname

\title{A VORTICITY-MAGNETIC FIELD DYNAMO INSTABILITY}

\author{Eric G. Blackman}
%\address{Institute of Astronomy, Madingley Road, Cambridge CB3 OHA, 
% England}              
\affil{Institute of Astronomy, Madingley Road, Cambridge CB3 OHA, 
England}

\author{Tom Chou}
%\address{DAMTP, University of Cambridge, 
%Cambridge, CB3 9EW, England}              
\affil{DAMTP, University of Cambridge, Cambridge, 
CB3 9EW, England}

\date{\today}

%\maketitle

%\vspace{3mm}
\bigskip
\begin{abstract}

We generalize the mean field magnetic dynamo to include local
evolution of the mean vorticity in addition to the mean magnetic
field.  The coupled equations exhibit a general mean field dynamo
instability that enables the transfer of turbulent energy to the
magnetic field and vorticity on larger scales.  The growth of the
vorticity and magnetic field both require helical turbulence which
can be supplied by an underlying global rotation.  The dynamo
coefficients are derived including the backreaction from the mean
magnetic field to lowest order. We find that a mean vorticity field
can actually seed exponential growth of mean magnetic field from
only a small scale seed magnetic field, without the need for a seed
mean magnetic field. The equations decouple when the fluctuating
velocity and magnetic field cross-correlations vanish, resulting in
the separate mean field magnetic and mean field vorticity dynamo
equations.

\end{abstract}
\bigskip
\ni Subject Headings:  magnetic fields: MHD; instabilities; accretion
disks; sun: magnetic fields; galaxies: jets.

%\vfill
%\eject
%]

%\keywords{}

%\pacs{95.30.Q, 47.65, 52.35.R, 91.25.C}

\section{Introduction}

The mean-field magnetic dynamo (MFMD) has been a standard
framework for understanding the origin of large scale
magnetic fields (B-fields)  in planets stars and galaxies
(\cite{MOF78,PAR79,RUZ88,BEC96}).  In this mechanism, a large
scale seed field grows exponentially at the expense of small
scale turbulent energy.  This can be distinguished from fast
dynamo turbulent amplification of B-fields (e.g. 
\cite{PAR79}), which occurs on the scale of the input
turbulence.  Many systems such as planetary and solar spots
(\cite{PET96}), and accretion and galactic disks
(\cite{ABR92,KIT94a}) also show evidence for large scale vortex
structures.  Jets may also be related.
In the absence of B-fields, the mean vorticity
equation is similar to the mean magnetic induction equation
(\cite{KIT94a}) and can allow mean field  vorticity dynamo
(MFVD) growth.  When the mutual coupling of vorticity and
B-field is considered, the mean quantity whose time
evolution is of interest has 6 components: 3 components each
for the mean B-field and mean vorticity.  Growth arises in
this coupled mean field dynamo (CMFD).  

%Vortex lines, like B-field lines, respond to stretching.  In
%systems with large magnetic Reynolds numbers, vorticity
%growth must ultimately saturate from B-field growth
%(\cite{KIT94b}). Saturation would be slowest when the scale of
%fastest vorticity growth is larger than that of the mean
%B-field.  

We first derive the equations for mean and fluctuating
quantities and show that the evolution equations for the mean
vorticity and B-field are non-trivial only when the
turbulence is at least weakly inhomogeneous and anisotropic. 
By expanding the turbulent quantities to first order in the
time-varying mean velocity and mean B-field, we derive the
initial mutual dynamo growth for several sets of initial
conditions. When cross-correlations between turbulent
velocities and turbulent B-fields vanish, mutual field growth
is described by decoupled MFVD and MFMD equations. However,
when the cross correlations do not vanish, the 
time evolution of the mean fields is coupled.
As a result, we  find that a seed mean vorticity and small
scale B-field can allow growth of the mean B-field even when
the latter is initially zero, in contrast to the decoupled theory.

We include the mean B-field backreaction on the velocity flows to
lowest order, thus deriving corrections to the kinematic magnetic
and vorticity dynamo coefficients.  However, we do not compute these
corrections to higher order, nor do we  purport to offer a fully
developed non-linear theory of MHD turbulence.  We make explicit
assumptions/approximations, but emphasize that none of these are
beyond those used in standard mean-field dynamo treatments.  (In
many treatments, the same approximations are imposed without
explanation.) We are aware of the controversies of the non-linear
backreaction (e.g.
\cite{PID81,KUL92,FIE96,CAT94,POU76,BLA96,BRA96,SUB97}), but
inasmuch as the  ``$\a^2$''(e.g. \cite{MOF78}) or ``$\a-\Om$'' (e.g.
\cite{PAR79}) dynamos are at least a framework for the study of 
B-field generation and comparison with observations, so too is the
more generalized treatment of the  vorticity-magnetic field dynamo
developed herein.  

\section{Derivation of Coupled Mean-Field Equations}

Consider the Navier-Stokes equation for incompressible
flows in the presence of B-fields,

\begin{equation}
\begin{array}{r}
\partial_{t} {\bfv}={\bfv} \ts(\nb\ts\bfv)
-\nabla p_{\em{eff}}
 + \nu\nabla^{2}{\bf v}+{\bf b}\cdot\nabla{\bf b}
\\[13pt]
+
{\bf F}({\bf x}, t)+\nabla\phi,
\end{array}
\label{NS}
\end{equation}

\noindent where $\nu$ is a constant viscosity,
${\bf F}$ is a forcing function, 
$p_{\em{eff}} \equiv p+b^2/2+v^2/2$, with $p\equiv P/\rho$
and ${\bf b}\equiv {\bf B}/(4\pi \rho)^{1/2}$ is the 
constant density normalized B-field. 
The $\nabla\phi$ includes potential forces such as gravity. 
The vorticity equation,

\begin{equation}
\partial_{t} {\bfw} =\curl(\bfv\ts\bfw)+
\nu\nabla^{2}{\bfw}
+\curl(\bfb\cdot\nabla \bfb)+\curl{\bf F}({\bf x}, t),
\label{VORTICITY}
\end{equation}

\noindent where ${\bfw} \equiv \curl {\bfv}$,
is obtained by taking the curl of  (\ref{NS}).
The induction equation for ${\bf b}$ is
 
\begin{equation}
\partial_{t}{\bf b} =\curl(\bfv\ts\bfb) + 
\nu_M\nabla^{2}{\bf b},
\label{INDUCTION}
\end{equation}

\noindent where $\nu_M$ is a constant magnetic viscosity.

Next, decompose $\bfw,\ \bfv$, and $\bfb$
into mean (indicated by an overbar or $\lb\ \rb$) and
fluctuating (indicated by prime) components
$\bfv=\bbv_T+\bfvp$, ${\bfw}=\bbw_T+\bfwp$ and
$\bfb=\bbb+\bfbp$, respectively with $\bbv_T=\bbv+\bbv_c$.
Here, $\bbv_c$ is a constant global velocity whose curl is
$\bbw_c$, and $\bbv$ and $\bbw$ are the time dependent mean
components whose spatial scales of variation are smaller
spatial than that of $\bbw_c$.  The $\bbw_c$  supplies
helicity to the turbulence and ensures that any growth of
$\bbw$ represents a transfer of angular momentum from the
turbulence.
%, consistent with angular momentum conservation. 
(Alternatively, the helicity could be supplied by
${\bf F}({\bf x},t)$.)

We assume that derivatives with respect to $\bf x$ or $t$
obey $\partial_{t,{\bf x}}\lb { X_i X_j} \rb =
\lb\partial_{t,{\bf x}}({X_i X_j})\rb$ and $\lb
\bar{X}_{i}X_{j}^{\prime} \rb = 0$ (Reynolds relations
(\cite{RAD80})), where $X_i={\bar X}_i+X^{\prime}_i$ are
components of vector functions of $\bf x$ and $t$.  For
statistical ensemble means, these hold when the 
correlation time scales are short relative to the variation
times of mean quantities.  For the spatial mean,
defined by $\lb X_{i}({\bf x},t)\rb = |\zeta|^{-3} \int^{{\bf
x}+ L}_{{\bf x}-L} X_{i} ({\bf s},{\rm t}) {\rm d^3}{\bf s}$,
the relations hold when the averaging is taken over a large enough
scale, such that $l \ll |{\bf \zeta}| \ll L$, where $L\sim
{\bar b}/\nabla {\bar b}$, and $\ell \sim b'/\nabla b'\sim v'/\nabla v'$.  

Subtracting the mean of (\ref {NS}) from
itself, and assuming $\nabla\phi'=0$, gives 

\begin{equation}
\begin{array}{l}
d_{t} \bfvp = 
\lb\bfvp\cdot\nabla\bfvp\rb-\bfvp\cdot\nabla\bfvp-
\bbv\cdot\nabla\bfvp
-\bfvp\cdot\nabla\bbv_T 
- \\
\nabla p'-\nabla b'^2/2+ 
\nabla \lb b'^2\rb/2 
-\nabla( \bfbp\cdot \bbb)+
\bfbp\cdot\nabla\bfbp
- \\
\lb\bfbp\cdot\nabla\bfbp\rb +
\bfbp\cdot\nabla\bbb+\bbb\cdot\nabla\bfbp 
+{\bf F}'({\bf x}, t)+
\nu\nabla^{2}\bfvp,
\label{VPRIME}
\end{array}
\end{equation}

\noindent where $d_t\equiv \partial_t +\bbv_c\cdot\nabla$. 
The average of  (\ref{VORTICITY}) is

\begin{equation}
\begin{array}{r}
d_t \bbw =\curl \lb\bfvp\ts\bfwp\rb
+ \nu\nabla^{2}{\bar{\bfw}}_T
+\curl \lb{\bfbp}\cdot\nabla{\bfbp}\rb + \\
\bbw_c\cdot\nabla\bbv
+\bbw\cdot\nabla\bbv_c+\bbw_c\cdot\nabla\bbv_c,
\end{array}
\label{OMEGABAR}
\end{equation}

\noindent where we have neglected terms second order in
time-varying mean quantities.  Subtracting 
(\ref{OMEGABAR}) from  (\ref{VORTICITY}) gives

\begin{equation}
\begin{array}{l}
d_{t} \bfwp = \bfwp\cdot\nb\bbv_T-\bbv\cdot\bfwp+\bbw_T
\cdot\nb\bfvp-
\bfvp\cdot\nabla\bfwp \\
-\bfvp\cdot\nb\bbw_T +
\bfwp\cdot\nabla\bfvp -
\curl\lb \bfvp\ts\bfwp\rb + \nu\nabla^{2}{\bf
\omega}^{\prime}+ \\
\curl ({\bfbp}\cdot\nabla{\bbb})
+\curl ({\bbb}\cdot\nabla{\bfbp}) +
\curl(\bfbp\cdot\nabla\bfbp)-\\
\curl\lb\bfbp\cdot\nabla\bfbp\rb 
+\curl{\bf F}'({\bf x}, t).
\end{array}
\label{OMEGAPRIME}
\end{equation}

\noindent Similarly, the equation for the mean 
B-field, derived by
averaging  (\ref{INDUCTION}) is

\begin{equation}
d_{t}\bbb =\curl\lb\bfvp\ts\bfbp\rb+\bbb\cdot\nabla\bbv_c
+\nu_M\nabla^{2}\bar{\bf b}.
\label{BBAR}
\end{equation}

\noindent Subtracting  (\ref{BBAR}) from (\ref{INDUCTION})
yields the equation for the fluctuating B-field 

\begin{equation}
\begin{array}{r}
d_{t} \bfbp = 
\bfbp\cdot\nb\bbv_T-\bbv\cdot\nabla\bfbp
+\bbb\cdot\nb\bfvp-\bfvp\cdot\nb\bbb
+ \\
\bfbp\cdot\nabla\bfvp -\bfvp\cdot\nabla\bfbp
-\curl\lb\bfvp\ts\bfbp\rb 
+ \nu_M\nabla^{2}\bfbp.
\end{array}
\label{BPRIME}
\end{equation}

Growth of $\bbb$ or $\bbw$ requires the $\nabla\times$ terms in
(\ref{BBAR}) and (\ref{OMEGABAR}) to be non-vanishing.  However,
when the turbulence is strictly  homogeneous in $\bfvp$ and $\bfbp$,
these quantities vanish.  For purely isotropic turbulence, the term
$\lb \bfvp\times\bfbp\rb$ in  (\ref{BBAR}) vanishes since it is
the average of a vector, while $\curl \lb\bfvp\ts\bfwp\rb\propto
\curl \nabla\lb v'^2\rb$ and  $\curl \lb\bfbp\cdot\nabla
\bfbp\rb\propto \curl \nabla \lb b'^2\rb$ vanish from Reynolds rules
and incompressibility.  Thus, anisotropy and inhomogeneity must be
present for nontrivial time evolution of mean fields.

We expand the turbulent quantities on the right of Eqs. 
(\ref{OMEGABAR}) and (\ref{BBAR}) to linear order in {\it both}
$\bbb$ and $\bbv$ using the equations for the fluctuating fields
thus generalizing  the approach of (\cite{FIE96}) where only the
mean B-field was used.  To find the lowest order terms, we assume
weakly anisotropic inhomogeneous turbulence: Terms linear in the
time-varying mean quantities contribute, but their averaged 0$^{th}$
order coefficients, are taken to be isotropic and homogeneous (still
allowing for reflection asymmetry). Iterating the equations using
the formal solutions for the turbulent fields
$\bfbp(t)=\bfbp(t=0)+\int d_{t'}\bfbp dt'$ and
$\bfwp(t)=\bfwp(t=0)+\int d_{t'}\bfwp dt'$, and using times
appropriately chosen such that the correlations
$\lb\bfvp(t)\ts\bfwp(0)\rb= \lb\bfvp(t)\ts\bfbp(0)\rb \simeq 0$, we
obtain to first order in mean quantities

\begin{equation}
\begin{array}{r}
\lb\bfvp\ts\bfwp\rb^{(1)}
=\lb\bfvp^{(0)}(t)\ts\int^t_0d_{t'} 
\bfwp^{(1)} dt' \rb +  \\
\lb\int_0^t d_{t'} 
\bfvp^{(1)} dt' \ts \bfwp^{(0)}(t)\rb,
\end{array}
\label{VPWP}
\end{equation}

%\begin{equation}
%\lb\bfvp\ts\bfwp\rb^{(1)}
%=\lb\bfvp^{(0)}(t)\ts\int^t_0d_{t'} 
%\bfwp^{(1)} dt' \rb
%+\lb\int_0^t d_{t'} 
%\bfvp^{(1)} dt' \ts \bfwp^{(0)}(t)\rb,
%\label{VPWP}
%\end{equation}

\noindent with similar expressions for
$\lb\bfbp\cdot\nabla\bfbp\rb^{(1)}$ and $\lb
\bfvp\times\bfbp\rb^{(1)}$. The calculation of these averages
requires Eqs. (\ref{VPRIME}), (\ref{OMEGAPRIME}), and
(\ref{BPRIME}) for the time integrands.  Using 
(\ref{VPRIME}) also requires an expression for the pressure,
which arises in (\ref{VPWP}) and in
$\lb\bfvp\ts\bfbp\rb$ via the terms

\begin{equation}
\lb\bfwp^{(0)}(t)\ts\int_0^t \nabla p'^{(1)} dt' 
\rb \,\,\, \mbox{and} \,\,\,
\lb\bfbp^{(0)}(t)\ts\int_0^t \nabla p'^{(1)} dt'\rb.
\label{PTERMS}
\end{equation}

Using isotropy, homogeneity, and Reynolds rules, we now show that
remarkably, terms of the form (\ref{PTERMS}) vanish in the derivation
of the mean field equations to the order considered.  Consider the
pressure in the isoentropic energy equation (hereafter neglecting
microphysical fluid and magnetic viscosities) $D_t P \equiv d_t
P+(\bfvp+\bbv)\cdot\nabla P= -\gamma P\nabla\cdot{\bf v}$, where
$\gamma$ is the adiabatic index (\cite{BATCHELOR,BOYD}).  We have
assumed incompressibility ($\nabla\cdot{\bf v}\simeq 0$ and constant
$\rho$) above, but here we must allow $\nabla\cdot{\bf v} \neq 0$ for
the last term in the energy equation: Even though $\nabla\cdot{\bf v}$
is small, the sound speed is large ({\it i.e.} $\gamma$ is large, see
\cite{BATCHELOR}) such that the magnitude of $DP/Dt$ cannot be
ignored.  This is why the energy equation is normally not useful in
incompressible approaches.  But here it will be symmetry properties of
the energy equation that account for the absence of a pressure
contribution when used in (\ref{PTERMS}) and the seemingly offending
term will not contribute.  Taking $p'\equiv P'/\rho$ we obtain

\begin{equation}
\begin{array}{r}
d_t p' =\lb\bfvp\cdot\nabla 
p'\rb -\bfvp\cdot \nabla{\bar p}-\bbv\cdot\nabla p'-
\bfvp\cdot\nabla p'+ \\
\gamma \lb p'\nabla\cdot\bfvp\rb - \gamma p'\nabla\cdot\bbv
-\gamma {\bar p} \nabla\cdot\bfvp - 
\gamma p'\nabla\cdot\bfvp,
\end{array}
\label{PPRIME}
\end{equation}

\noindent where $\nabla\cdot\bbv_{c} \simeq 0$, but
$\nabla\cdot\bbv, \nabla\cdot\bfvp \neq 0$.  The averaged terms in
 (\ref{PPRIME}) make no contribution when inserted into  
(\ref{PTERMS}) according to the Reynolds rules.  We only consider
terms {\it explicitly} linear in time-varying mean quantities, and
iterate to first order in the correlation time; higher order terms
would come from successive iterations in the equations for the
fluctuating variables.  This  first order smoothing approximation
(FOSA)(\cite{KRA80}) is ``justified'' when the correlation time of
the turbulence is less than the eddy turnover time
(\cite{RUZ88,BLA95}), as is possibly the case in galaxies
(\cite{RUZ88}), or if higher order correlations are otherwise
reduced. Integrating  (\ref{PPRIME}), and using the assumptions
discussed above, 

\begin{eqnarray}
\nabla p'^{(1)} = \nabla p'^{(1)}(0)-\nabla\!
\int (\bfvp^{(0)}\!\cdot\!\nabla {\bar
p}+\bbv\!\cdot\!\nabla
p'^{(0)} \quad  \\ \nonumber 
+\ga p'^{(0)}\nabla\cdot\bbv+\ga
\bar{p}\nabla\cdot\bfvp^{(0)}) dt'.  
\end{eqnarray}

%\begin{equation}
%\nabla p'^{(1)} = \nabla p'^{(1)}(t=0) - \nabla
%\int (\bfvp^{(0)}\cdot\nabla {\bar p}+\bbv\cdot\nabla
%p'^{(0)}+\ga p'^{(0)}\nabla\cdot\bbv+\ga \bar{p}\nabla\cdot\bfvp^{(0)}.
%+\ga p'^{(0)}\nabla\cdot\bfvp) dt'.  
%\end{equation}

\noindent The pressure dependent contribution to $\lb \bv0\ts
\bw0\rb$ can then be written

\begin{equation}
\begin{array}{l}
\displaystyle \lb \bw0\ts \int_{0}^{t}d_t'\nabla p' dt' \rb_{k}
 = \ep_{ijk}\ep_{ims} \times
\quad \quad \quad \\[13pt] 
\displaystyle \quad\quad \displaystyle \int_{0}^{t}\!dt'\int_{0}^{t'}\!dt'' 
\lb (\partial_{j} {\bar v}_l \partial_l p'^{(0)}+ 
{\bar v}_l \partial_j \partial_l p'^{(0)}+ \quad  \\[13pt] 
\displaystyle \quad\quad \partial_j v'^{(0)}_l 
\partial_l{\bar p}+v'^{(0)}_l\partial_j\partial_l{\bar p}
+ \ga p'^{(0)}\partial_j\partial_l\bar{v}_l+\\[13pt]
\displaystyle\hspace{1cm} \ga\partial_j p'^{(0)}\partial_l\bar{v}_l+
\ga{\bar p}\partial_j\partial_l v'^{(0)}_l+
\ga\partial_j{\bar p}\partial_l v'^{(0)}_l) \\[13pt]
\quad \hspace{3cm}\times  (\partial_m v'^{(0)}_s-
\partial_s v'^{(0)}_m)\rb \quad  \\[13pt] 
\displaystyle \quad =  
{1 \over 6}\int_{0}^{t}dt' \int_{0}^{t'} dt''
\lb \bw0(t')\cdot\bw0(t'') \rb \partial_{k}\bar{p}(t),
\end{array}
\label{P3}
\end{equation}

%\begin{equation}
%\begin{array}{ll}
%\displaystyle \lb \bw0\ts \int_{0}^{t}d_t'\nabla p' dt' \rb_{k}
%& = \displaystyle \ep_{ijk}\ep_{ims}\int_{0}^{t}dt' 
%\int_{0}^{t'}dt'' 
%\lb (\partial_{j} {\bar v}_l \partial_l p'^{(0)}+ 
%{\bar v}_l \partial_j \partial_l p'^{(0)}+ \partial_j v'^{(0)}_l 
%\partial_l{\bar p}+v'^{(0)}_l\partial_j\partial_l{\bar p} \\
%\displaystyle \: & \displaystyle\hspace{-2cm} + 
%\ga p'^{(0)}\partial_j\partial_l\bar{v}_l+
%\ga\partial_j p'^{(0)}\partial_l\bar{v}_l+
%\ga{\bar p}\partial_j\partial_l v'^{(0)}_l+
%\ga\partial_j{\bar p}\partial_l v'^{(0)}_l)
%(\partial_m v'^{(0)}_s-\partial_s v'^{(0)}_m)\rb \\
%\displaystyle \: & \displaystyle =  
%{1 \over 6}\int_{0}^{t}dt' \int_{0}^{t'} dt''
%\lb \bw0(t')\cdot\bw0(t'') \rb \partial_{k}\bar{p}(t),
%\end{array}
%\label{P3}
%\end{equation}

\noindent where we assume mean fields vary on time
scales longer than the fluctuating fields, and $\nabla
p'^{(1)}(t=0)$ is uncorrelated with $\bfbp(t)$ or ${\bfw}'(t)$. 
The only surviving term arises from the third term on the right hand
side of  (\ref{P3}) and the vanishing of the remaining terms
follows from careful application of isotropy ({\it i.e.} rank 2 and
rank 3 averaged tensors of fluctuating 0$^{th}$ order quantities are
proportional to $\delta_{ij}$ and $\ep_{ijk}$ respectively) and
homogeneity ({\it i.e.} $\partial_{i}\langle X_{j}X_{k}\rangle^{(0)}
= 0$) of the 0$^{th}$ order turbulence, {\it without} invoking
$\nabla\cdot\bfvp=0$. The surviving term in (\ref{P3}) vanishes when
put inside the curl in  (\ref{OMEGABAR}).  A similar analysis
holds for the second term in (\ref{PTERMS}) when put into
(\ref{BBAR}); thus, the pressure does not contribute to
(\ref{OMEGABAR}) or (\ref{BBAR}).

Approximating time integrals 
%$\lb\bfwp\ts\bfbp\rb^{(1)},\, \lb\bfbp\cdot\nabla
%\bfbp\rb^{(1)}$ and $\lb\bfvp\times\bfbp\rb^{(1)}$ 
by factors of the correlation time $\tau_{c}$ (\cite{RUZ88}), and freely
employing Reynolds rules and incompressibility we obtain

\begin{equation}
\begin{array}{l}
\displaystyle (3/\tau_{c})
\curl\lb\bfvp\ts\bfwp\rb^{(1)}
= \displaystyle \lb \v0_{i}\w0_{i}\rb(\curl\bbw) + \\
\lb\v0_{i}\v0_{i}\rb
\nabla^{2}{\bbw} +
2\lb\w0_{i}\b0_{i}\rb\nabla^{2}\bbb  + \\
\lb\epsilon_{ijk}\w0_{k}\partial_{i}
\b0_{j}\rb(\curl\bbb)
-\lb\v0_{i}\b0_{i}\rb\nabla^{2}(\curl\bbb)
\label{MESS1}
\end{array}
\end{equation}

\noindent with similar expressions for
$\curl\lb\bfbp\cdot\nabla\bfbp\rb^{(1)}$ and $\curl\lb
\bfvp\times\bfbp\rb^{(1)}$.  Upon substituting these into
(\ref{OMEGABAR}) and (\ref{BBAR}), the curls can be pulled onto
the  $\bbw$ and $\bbb$ from homogeneity of 0$^{th}$ order
averages.  Note that $\bbw_c$ and $\bbv_c$ contribute to the
0$^{th}$ order quantities, and thus do not show up explicitly in
(13). After some simplification, we obtain

\begin{equation}
\begin{array}{r}
d_{t}\bbw = \alpha_{0}(\curl \bbw) +
\alpha_{1}(\curl\bbb) +\beta_{0}\nabla^{2}\bbw
+\beta_{1}\nabla^{2}\bbb \\
- \alpha_{2}\nabla^{2}(\curl\bbb)
+\bbw_c\cdot\nabla\bbv
+\bbw\cdot\nabla\bbv_c+\bbw_c\cdot\nabla\bbv_c,
\end{array}
\label{WEQM}
\end{equation}

\noindent and

\begin{equation}
d_{t}\bbb = \alpha_{2}(\curl\bbw)+
\alpha_{3}(\curl\bbb)+\beta_{2}\nabla^{2}\bbb+
\bbb\cdot\nabla\bbv_c
\label{BEQM}
\end{equation}

\noindent where the coefficients are:

\begin{equation}
\begin{array}{l}
\alpha_{0} = (\tau_{c}/3)\langle\bw0\cdot\bv0 \rb \\
\alpha_{1} = (\tau_{c}/3)\langle\bw0\cdot\curl\bb0\rangle \\
\alpha_{2} = (2\tau_{c}/3)\langle\bb0\cdot\bv0\rangle \\
%\alpha_{3} =(\tau_{c}/3)\langle\bb0\cdot\curl\bb0\rangle-\alpha_{0} \\
\alpha_{3} =(2\tau_{c}/3)\langle\bb0\cdot\curl\bb0\rangle-\alpha_{0} \\
\beta_{0} =(\tau_{c}/3)\langle\bv0\cdot\bv0+\bb0\cdot\bb0\rangle \\
\beta_{1} = (2\tau_{c}/3)\lb\bw0\cdot\bb0\rb \\
%\beta_{2} =(\tau_{c}/3)\langle\bv0\cdot\bv0+\bb0\cdot\bb0\rangle,
\beta_{2} =(\tau_{c}/3)\langle\bv0\cdot\bv0+2\bb0\cdot\bb0\rangle,
\end{array}
\label{COEF}
\end{equation}

\section{Solutions and Discussion}

The coefficients $\alpha_{3}$ and $\be_2$ in (\ref{BEQM})  are
modified helicity and diffusion coefficients of the standard
magnetic dynamo theory.  A similar, but not identical, set of
helicity and diffusion coefficients enter (\ref{WEQM}) for the
mean vorticity.  The cross-correlations $\a_1,\a_2,\be_1$, taken
to vanish in \cite{VAI83}, need not vanish, as they preserve all
MHD symmetries in (\ref{WEQM}) and (\ref{BEQM}). 
We will now see how the finite helicity is required for both
B-field (\cite{MOF78}) and vorticity growth (\cite{KIT94a}).  

To show the growth, we assume constant dynamo coefficients and
consider the simple case when the last 3 terms in (\ref{WEQM}) and
the last term in (\ref{BEQM}) are negligible.  This assumption is
valid in systems for which $\bbw_c$ is perpendicular to the gradient
of the time varying mean quantities, and for which the former is
relatively uniform or perpendicular to the dominant B-field
component.  (The influence of $\bbw_c$ is always present however, in
providing reflection asymmetry to the 0$^{th}$ order turbulent
quantities.) The simplified equations yield a generalized
vorticity-magnetic $\a^2$ dynamo.  

The mean-field equations are then solved by assuming
solutions of the form exp$(\sigma t -i\vec{k}\cdot\vec{r})$ to
obtain the linear algebraic equations $\sigma \vec{X} = {\bf
M}\vec{X}\, , \vec{X} = (\bar{\omega}_{x}, \bar{\omega}_{y},
\bar{\omega}_{z}, \bar{b}_{x}, \bar{b}_{y}, \bar{b}_{z})$. 
%Alternatively, we can take the curl of (\ref{BEQM}) and work
%with the current $\bar{\bf j} = \curl\bbb$. 
The eigenvalues
$\sigma$ of the dynamical matrix ${\bf M}(\vec{k})$ determine
the time evolution of the fields at wavevector $\vec{k}$.  
Different behavior can be inferred from the general solution
$\vec{X}(\vec{r}, t) = \sum_{j=1, 6}c_{j}\vec{\xi}_{j}\,
exp(\sigma_{j}t - i\vec{k}\cdot\vec{r})$.  Here, $c_{j}$ are
depend on the initial values ${X_j}(t=0)$, and $\vec{\xi}_{i}$
are eigenvectors which are linear combinations of $\bbw$ and
$\bbb$.  When cross-correlations between the turbulent
velocities and B-fields are non-vanishing, the mean vorticity
and B-fields are naturally coupled.  The general $6\times 6$
matrix equation can be solved numerically for $X_j$ and contains
all decaying and growing modes.  Because the $c_j$ are
combinations of the initial $X_j(0)$, for growing modes, all
components of $\bar{\bf b}$ and $\bar{\bfw}$ will in general
grow unless very specific initial conditions ${\bf X}(t=0)$ and
values of $k$ and $\sigma(k)$ make certain $c_{j}$ vanish.  

First we consider thin disk systems which have only in-plane mean
field components (while allowing the turbulence to be 3-D), {\it
i.e.,} $\bar{\bf v}  = ({\bar v}_{x}, {\bar v}_{y}, 0),\, \bar{\bf
b} = ({\bar b}_{x}, {\bar b}_{y}, 0)$. The three possible poles are
$\sigma_{0} = -\beta_{0}k^2$ and $2\sigma_{\pm} =
-(\beta_{0}+\beta_{2})k^{2} \pm \vert
k\vert\sqrt{4\alpha_{1}\alpha_{2}+
(\beta_{0}-\beta_{2})^{2}k^{2}+4\alpha_{2}^{2}k^{2}}$.  which have
growing modes ($\sigma > 0$) when $k^{2} < \alpha_{1}\alpha_{2}/
(\beta_{0}\beta_{2} - \alpha_{2}^{2})$ (if and only if
$\alpha_{1}\alpha_{2}/ (\beta_{0}\beta_{2} -\alpha_{2}^{2}) > 0$). 
Thus, the growth of $\bar{\omega}_{z}$ and $(\bar{b}_{x},
\bar{b}_{y})$ depends on the signs and magnitudes of $\alpha_{1},
\alpha_{2}, \beta_{0}$, and $\beta_{2}$.  %anti-dynamo theorem
because we allow the small scale %fields to be three dimensional.  

The fully 3-D case requires numerical calculation of
$\sigma_{i}$, but we can obtain approximate behavior by
truncating (\ref{WEQM}) and (\ref{BEQM}) to lowest order in mean
field gradients.  By taking the curl of (\ref{BEQM}) and using
$\bar{\bf j}$ we see that terms with one higher derivative in
the mean quantities are reduced by a factor $\ell/L$.  If
$\alpha_{2} \propto \lb\bv0\cdot\bb0\rb$ is negligible
initially, it will remain so (\cite{CHA97,KRAICHNAN}). 
Neglecting only the $\alpha_{2}$ term in  (\ref{WEQM}) and
(\ref{BEQM}), the six roots are $\sigma = -\beta_{0}k^{2},
-\beta_{2}k^{2}, -\beta_{0}k^2\pm\alpha_{0}\vert k \vert,
-\beta_{2}k^{2}\pm \alpha_{3}\vert k\vert$, which displays two
growing modes for small $k$ depending on the signs of
$\alpha_{0}$ and $\alpha_{3}\, (\beta_{0},\, \beta_{3} > 0)$.

Note that if cross-correlations between turbulent velocities and
magnetic fields vanish, the equations for $\bbw$ and
$\bar{\bf b}$ decouple into separate MFMD and MFVD equations of
the form $\partial_{t}{\bbm} =\a \curl\bbm+ \beta \nabla^2\bbm$
where $\bar{\bf m} = \bbw$ or $\bar{\bf b}$, with their
respective helicity and diffusion coefficients.  The time
evolution of the decoupled dynamos is determined by the
eigenvalues $\sigma = -\beta k^2,\, \pm \alpha \vert k\vert -
\beta k^2$. Thus, there is a growing solution $(\sigma = \vert
\alpha \vert \vert k\vert - \beta k^2 > 0)$ for small
wavevectors $\vert k \vert < \vert \alpha\vert /\beta$. The most
unstable mode is $k^{*} = \vert \alpha \vert /(2\beta)$ which
grows at a rate $\sigma(k^{*}) = \vert \alpha^2 \vert/(4\beta
)$, like the usual $\a^2$ dynamo (\cite{MOF78}).

To conclude: We have generalized the mean-field dynamo to include
mean vorticity and have considered solutions of an $\a^2$
vorticity-magnetic dynamo.  An important implication of (\ref{WEQM})
and (\ref{BEQM}) is that the mean B-field can grow even when its
initial value is zero because a seed mean B-field can be generated
by a coupling between the small scale B-field to the small scale
velocity, and the latter to the mean vorticity.  This feature should
be preserved in more general treatments as it simply results from
the coupling between  (\ref{WEQM}) and (\ref{BEQM}) and only
requires non-vanishing cross correlations.  Growth requires helical
turbulence which can be supplied by an underlying global rotation. 
Astrophysical applications of more extensive treatments may include
accretion disks, where vortices and B-fields can combine to
concentrate emerging radiation into strong beams
(\cite{ACO90,ABR92}).  Also, Yoshizawa \& Yokoi (1993) showed that
an equation similar to (\ref{BEQM}) can lead to generation of
toroidal fields whose upward pressure drives a jet that subsequently
may develop poloidal fields.  The CMFD may provide the sustaining
feedback.

\end{document}